# Electron gas polarization effect induced by heavy H-like ions of moderate velocities channeled in a silicon crystal

E. Testa[1], D. Dauvergne[1*], A. Bräuning-Demian[2], F. Bosch[2], H. Bräuning[3], M. Chevallier[1], C. Cohen[4], A. Gumberidze[2], S. Hagmann[2], A. L'Hoir[4], R. Kirsch[1], C. Kozhuharov[2], D. Liesen[2], P.H. Mokler[2], J.-C. Poizat[1], C. Ray[1], J.-P. Rozet[4], Th. Stöhlker[2], S. Toleikis[2], M. Toulemonde[5] and P. Verma[2]

[1] Institut de Physique Nucléaire de Lyon, CNRS- IN2P3 and Université Claude Bernard Lyon 1, F-69622 Villeurbanne, France

[2] Gesellschaft für Schwerionen Forschung (GSI), D-64291 Darmstadt, Germany

[3] Institut für Kernphysik, Justus Liebig Universität, D-35392 Giessen, Germany

[4] Institut des Nano-Sciences de Paris, CNRS-UMR75-88, Universités Paris VI et Paris VII, 75251 Paris cedex 05, France

[5] Centre Interdisciplinaire de Recherche Ions-Lasers, UMR 11 CEA-CNRS, 14040 Caen cedex, France

**Abstract :**

We report on the observation of a strong perturbation of the electron gas induced by 20 MeV/u $U^{91+}$ ions and 13 MeV/u $Pb^{81+}$ ions channeled in silicon crystals. This collective response (wake effect) induces a shift of the continuum energy level by more than 100 eV, which is observed by means of Radiative Electron Capture into the K and L-shells of the projectiles. We also observe an increase of the REC probability by 20-50% relative to the probability in a non-perturbed electron gas. The energy shift is in agreement with calculations using the linear response theory, whereas the local electron density enhancement is much smaller than predicted by the same model. This shows that, for the small values of the adiabaticity parameter achieved in our experiments, the density fluctuations are not strongly localized at the vicinity of the heavy ions.

---

[*] Corresponding author: d.dauvergne@ipnl.in2p3.fr . Tel.: (33) 4 72 44 62 57 . Fax: (33) 4 72 43 14 52

**Introduction**

Highly charged ions of moderate velocities induce a strong perturbation of the electron gas in a solid. This perturbation consists in a dynamic screening of the projectile charge by the electrons of the medium. If the particle propagates much faster than the Fermi velocity of the electrons, a trailing polarization cone takes place, which is due to the collective excitation of the electron gas (wake effect). The wake potential manifests itself through the electric field, acting as the stopping force on the projectile, and is also responsible for a Stark splitting of bound states of an ion, which influences the transport of excited states in solids. For a review on the wake effect, see e.g. refs. [1,2,3] and references therein. In the vicinity of the projectile, the polarization of the electron gas produces a shift, relative to vacuum, of the continuum energy levels in the projectile frame. This reduces the absolute binding energies of electrons in an ion. Echenique *et al.*[2] estimated this shift, using the linear response theory, to be $\lim_{r \to 0} \left( \Phi - Z_p e^2 / r \right) = -\dfrac{\pi Z_p \hbar \omega_p}{2 v / v_0}$ , where $\Phi$ is the potential, $Z_p$ and $v$ are the ion charge and velocity, respectively, $\hbar \omega_p$ the plasmon excitation energy of the target electrons, and $v_0$ the Bohr velocity. For a free electron gas of density $\rho_e$, $\omega_p = \left( 4\pi \rho_e e^2 / m_e \right)^{1/2}$, where $m_e$ is the electron mass. Also related to the wake are the fluctuations of the electron density around the projectile. Still within the first order perturbation, the authors of ref.[2] also predicted that the relative enhancement of the local electron density $\rho_e$ should be $\Delta \rho_e / \rho_e = \pi Z_p v_0 / v$.

Practically, transitions between bound states of an ionic projectile may be affected only by Stark splitting of these states, whereas transitions between continuum states and bound states are modified in energy. Among them, Radiative Electron Capture (REC) of target electrons provides a local probe for the electron gas polarization in a solid. First, since REC consists in the emission of a photon, the energy of which is the sum of the kinetic energy of the target electron in the ion frame, and the binding energy in the final bound state, it gives access to the energy shift (decrease of the emitted photon energy) [3,4]. Second, REC into deeply bound states of a heavy ion is a very localized process (at the scale of the final bound orbital size). So the measurement of absolute REC probabilities is a test for a possible enhancement of the local electron density at the projectile site.

The problem arising is that strong enough perturbations to be measurable are obtained for high $Z_p/v$ values, for which REC is hardly observable in solids, because it requires inner-shell vacancies.

Ion channeling leads to a non homogeneous flux inside a crystal, preventing ions from undergoing close impact parameter collisions with target atoms. Thus channeled ions sample mainly the quasi-free valence electron gas, and close interaction with core electrons is substantially attenuated. This allows REC to be the dominant electron capture process, even at low energy. Another interesting feature related to the impact parameter distribution is that detailed information can be obtained by analyzing the shape of REC lines, which depends on the longitudinal momentum distribution of the target electrons in their initial state (Compton profile). In particular this enables to identify the contributions of core electrons and valence or conduction electrons [6]. Finally, the knowledge of the ion flux in the crystal, combined with the precise measurements of the REC line intensities, allows one to evaluate the electron density at the ion site.

**Experiment**

We used 20 MeV/u $U^{91+}$ and 13 MeV/u $Pb^{81+}$ ions extracted from the GSI-ESR storage ring, for which $Z_p v_0/v$ values are 3.26 and 3.57, respectively. The process for cooling, deceleration and slow extraction of H-like ions by radiative recombination inside the electron cooler has been described in ref. [7]. During the extraction cycle, a continuous beam of some $10^4$ ions per second is sent onto the target with an angular divergence suitable for channeling experiments. The beam impact on the target is less than 3 mm in width, and 7 mm in height. A 9.6 µm thick (111) Si crystal (tilted at 35° for alignment along the <110> axis) was used as a target during the experiment with $U^{91+}$ ions. For the $Pb^{81+}$ ion experiment, a thin (0.8 µm) (100) silicon crystal was tilted at 45° to allow the same axial orientation. Transmitted ions were charge- and energy-analyzed by a magnetic spectrometer, and detected at the focal point by a 2D- position sensitive particle detector.

X-rays emitted at the target were detected by a 1 cm thick germanium detector at 90° from the beam direction. The detector was set at 135 mm from the beam impact in both experiments. The Doppler broadening of X-rays by the detector angular aperture was limited by vertical collimating slits of lead (6 mm) in the $U^{91+}$ experiment, and of tantalum (8 mm) for the $Pb^{81+}$ experiment.

The acquisition was done event by event, allowing for instance the selection of coincidences between X-rays and a given charge state at emergence.

**Results**

Part of the experimental details and results has been reported already in ref. [8] for the $U^{91+}$ experiment. In particular the charge state distribution for <110> axial orientation of the 11 μm thick target showed a fraction of ~25% of frozen ions, and about the same for ions emerging as 90+. For the latter most of the capture was due to REC. An illustration is given in fig. 1, which shows X-rays recorded for both axial and random incidences during this experiment. The axial spectrum is recorded in coincidence with ions emerging from the crystal with the charge 90+, i.e. ions having captured only one electron. In the axial spectrum, the K-REC and L-REC peaks are observed with a very good statistics, whereas they are reduced by nearly two orders of magnitude for a random orientation. The non-radiative capture (or Mechanical Electron Capture, MEC) is the dominant capture process in random geometry. In this case, charge state equilibrium is reached and K- and L-shell vacancies are rapidly filled below the surface. The signature of MEC is still present in the axial spectrum, with the presence of K and L lines, that come from a large part from decays after MEC into excited states ($n>2$). For well channeled ions, like ions emerging as 90+, most of the MEC events occur in the thin amorphous layers at the crystal surfaces. For a more detailed discussion on the X-ray spectra, see ref. [8].

The situation is similar for 13 MeV/u $Pb^{81+}$ ions channeled in a 1.1 μm crystal. In this case the frozen ion fraction was 60%, and the 80+ fraction was 20%. The target being much thinner, the statistics on REC are lower than in the former experiment. A spectrum recorded for the axial orientation, in coincidence with He-like transmitted ions, is shown in Fig. 2. The proportion of K lines is higher, due to the larger relative contribution of amorphous layers to the MEC single capture events.

Both spectra on Figs 1 and 2 are not corrected for detection efficiency.

The measurements of absolute energy shifts for REC peaks require precise energy calibration. This was made in various ways in the whole energy range described here. First radioactive γ-ray sources were used, without the beam on target. We also used the Pb and Ta Kα fluorescence lines (for instance the Pb Kα lines can be observed on the spectra of fig. 1). Additional information for the linear calibration offset is obtained by setting the differences between K-and L-REC peak energies equal to the Kα ener-gies. The precise knowledge of the K and L peak energies in the laboratory frame allows the correction of the Doppler shift of the photon energy emitted by fast projectiles: the derivative of the Doppler shift is maximum at 90°, and so a small deviation of the mean laboratory angle has to be taken into account. The energy loss inside the target was taken into account to determine the mean energy of REC photons emitted during the path of the ions in the crystal. For $Pb^{81+}$ in the 1.1 µm thick target the energy loss was almost negligible.

As done in ref. [6], we have performed simulations based on the statistical flux equilibrium for channeled ions in the crystal, according to the experimental conditions (beam angular divergence, energy loss, Doppler shift and broadening, detector resolution). These simulations provide the full calculation of the REC line shape, conditioned by the local electron density sampled by channeled ions. The Compton profile is calculated using the local density approximation of a free valence electron gas. For core electron, an impact parameter dependent Fourier transform of spatial wave functions is performed. Figure 3 shows the results of such a simulation for the K-REC line shape of $U^{91+}$ ions in the laboratory frame. This simulation was made assuming that all incident ions in channeling conditions could undergo REC in the crystal. Among them, ions with a high transverse energy can approach the target atoms sufficiently close, so that they can capture core electrons of silicon. The corresponding Compton profile is much broader than the Compton profile of valence electrons. As silicon atoms have 10 core electrons and 4 valence electrons, one can see that the calculated rate for the capture of core electrons is strongly reduced by channeling. In order to reproduce our experimental observations (angular scans), we used a beam angular divergence made of two components: a narrow one (85% of the beam), of rms $\sigma_x=\sigma_y=0.2$ mrad, and a broad one (15%), rms $\sigma_x=\sigma_y=3.5$ mrad (i.e. larger than the channeling critical angle). According to this, the mean sampled electron densities are calculated to be 0.16 $e^-Å^{-3}$ for valence electrons, and 0.053 $e^-Å^{-3}$ for core electrons. The experimental peak is superimposed on the calculations. One can see that the calculated core electron contribution is obviously overestimated, and that valence electrons represent almost all the contribution to the K-REC peak. Actually, as we already stated in ref. [8], ions with high enough transverse energy to approach the target atoms as close as 0.5 Å undergo many MEC capture events, which rapidly fill their inner-shell and prevent REC to occur. So only ions with a restricted transverse energy are able to make REC in the target. The simulations show that their mean sampled electron density is 0.17 $e^-Å^{-3}$. This is an important point since, as calculated for instance in ref. [4], the strong dependence of REC cross section on the relative velocity of tar-get electron would make REC peaks strongly asymmetric, which induces a negative average energy shift of the REC peak because of the Compton profile. This has to be taken into account to correctly estimate the wake shift. In the present case where mainly valence electrons are captured, the shift due to the Compton profile is minimum. The comparison between calculated and measured peaks in Fig.3 exhibits a shift by about -100 eV for the experimental one, that we attribute to the wake effect.

The shift for the L-REC lines, which has to be the same as for K-REC, is also measured, and is found in agreement within the calibration uncertainties. Thus we present in table 1 the measured values in both experiments, which are averaged for K- and L-REC. These shifts were measured by selecting REC photons in coincidence with He-like emergent ions, i.e. ions with a small transverse energy. The advantages are: i) contributions of core electrons to REC are even smaller than in Fig.3, ii) their energy loss is minimum, iii) no additional shift on the L-REC energy comes from the various lower charge states. The error bars come from the determination of the peak positions in the spectra, the X-ray energy calibration, the evaluation of the energy loss in the case of the uranium experiment, and the theoretical knowledge of the binding energies for the He-like ions (taken from ref.[9]). Calculations for the energy shift were made using the commonly admitted formula indicated in the introduction, and using an electron gas density equal to the average valence electron density of silicon (i.e. $\hbar\omega_p = 16.6$ eV). This value assumes that the whole valence electron gas contributes collectively to the energy shift, i.e. that the polarization is a long range effect. It also supposes that core electrons of silicon do not contribute significantly to the collective response, which may somehow underestimate this response. The agreement between experimental and theoretical values is quite good.

As for the absolute REC cross sections, they can be evaluated, although the absolute detector efficiency was not measured by another mean than by geometrical calculations. When looking at figs 1 and 2, one can get an absolute normalization of the X-ray yields, since all the ions having captured at least one electron in the crystal have emitted one - and only one – photon corresponding to the filling of the K-shell. These photons are $K_{\alpha\beta...}$ and K-REC (in this case the fluorescence yield is 100% for the initial K-vacancy). For a given charge state used to select X-rays, the yields of K- and L-REC photons in the spectra (corrected from the intrinsic detector efficiency), multiplied by the charge fraction, provide the K- and L-REC probabilities. Simulations, to be described in a forecoming paper, allow us to estimate the mean unperturbed electron density $\langle \rho_e \rangle$ sampled by channeled ions, as a function of their emerging charge state. Accounting for the angular distribution of REC photons in the laboratory frame [10], one can then give the absolute value of the K- and L-REC probabilities. We use the Stobbe formulae [11] of REC cross sections $\sigma_{REC}$, which are commonly used to estimate the absolute REC cross sections as a function of the adiabaticity parameter $\eta = (v/Z_p v_0)^2$ (assuming non relativistic velocities for both the projectile and the electron in a bound orbital, $\eta$ is the ratio of the kinetic energy of a target electron viewed by the projectile to the binding energy in the final state) [10]. The K- and L-REC probabilities are found to lie between 20% and 60% above the values given by $P(REC) = 1 - \exp(-\sigma_{REC} \times \langle \rho_e \rangle \times \ell)$, where $\ell$ is the crystal thickness. Moreover, the non-relativistic dipole approximation calculations of Stobbe formulae tend to be systematically above the experimental values by at least 25%, as reviewed in ref. [10]. Note that exact relativistic calculations have been performed by J. Eichler and A. Ichihara in the case of 20 MeV/u $U^{91+}$ ions [12]. Their K-REC cross section is 20% lower than Stobbe's one, and higher n-REC cross sections are in agreement within 10%. This justifies the use of Stobbe's values as a fairly good reference on one hand, and, on the other hand, makes our electron density enhancement even higher. Actually, some solid state measurements at high $Z_p/v$ values tend to lead to higher values of the REC cross sections. For instance, our former measurements of K-REC with 60 MeV/u $Kr^{36+}$ ions ($Z_p v_0/v = 1.1$) under channeling conditions agree perfectly with the Stobbe formula (with an absolute uncertainty of $\pm 10\%$) [6]. Tribedi et al. [5] used the linear response scaling [2] of the local electron density to explain the excess of REC cross sections in channeling relative to gas targets. We prefer to consider an increase of electron density instead of an increase of cross sections, which are defined for a single ion-electron (or ion-atom) collision. In the present case, this density enhancement would be by a factor 10, which is certainly not observed. However, our values are significantly above the calculations using a non-perturbed electron gas. We can already claim that the local density enhancement by the wake effect is not as localized as predicted by the linear response theory (which could be expected since we are not at all in a regime where $Z_p v_0/v << 1$). On the other hand, the induced potential at the projectile site results from an integration of the polarization over a very large scale (a typical scaling of the perturbation is given by $v/\omega_p \sim 10$Å), which makes the perturbation approximation more realistic for the REC energy shift than for the REC probability. However, in order to get a better understanding of the whole dynamic polarization effect on REC energies and probabilities, the comparison with REC during collisions in gas is highly desirable, as well as a non perturbative theoretical description of the effect.

**Summary and conclusion**

We have measured K- and L-REC with highly-charged, decelerated heavy ions in channeling conditions, which allows to keep inner-shell vacancies, and to perform ion-electron interaction inside a solid target. We report on a very strong value of the energy shift due to the polarization of the dense electron gas sampled in the solid (~100 eV), in good agreement with theoretical expectations based on first order perturbation. Our data show an enhancement of the total REC probabilities with respect both to theoretical calculations (by 20-60% relative to the non-relativistic dipole approximation cross sections) and to measurements performed at $Z_p v_0/v$ ~1 (by more than 50% compared to gas target experiments). This is likely due to the polarization effect around an ion, this effect being much smaller than predicted by the linear response theory.

We would like to thank J. M. Pitarke and M. Seliger for helpful discussions, and J. Eichler and A. Ichihara for providing us with REC cross sections. The support by GSI-IN2P3 collaboration agreement # 97-35 is acknowledged.

|  | $Z_p v_0/v$ | ΔE REC (eV) | $-\dfrac{\pi Z_p \hbar \omega_p}{2v/v_0}$ (eV) |
|---|---|---|---|
| $U^{91+}$ 20 MeV/u | 3.26 | $-92 \pm 41$ | $-85$ |
| $Pb^{81+}$ 13 MeV/u | 3.57 | $-122 \pm 47$ | $-93$ |

Table 1 : measured and calculated energy shifts of the continuum due to the electron gas polarization effect. The calculated values are given for a plasmon energy of 16.6 eV.

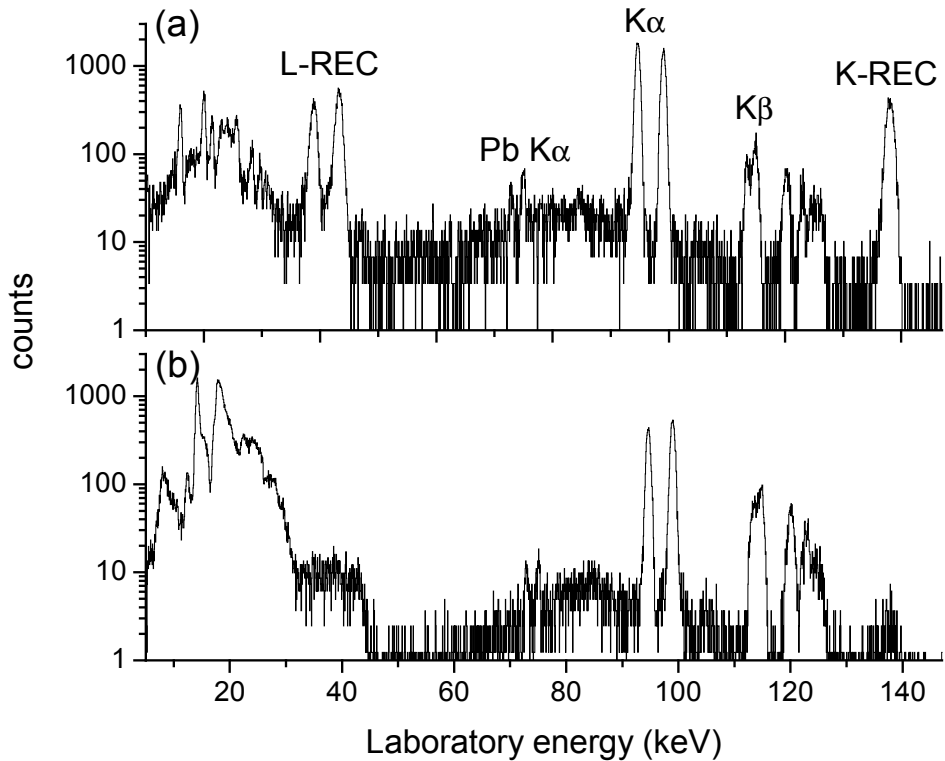

Figure 1: X-ray spectra detected at 90° from the beam direction for $U^{91+}$ ions incident on a 11 μm thick silicon crystal. (a) <110> axial orientation, in coincidence with ions transmitted in the 90+ charge state. (b) random orientation. Both spectra are normalized to the same number of selected transmitted ions.

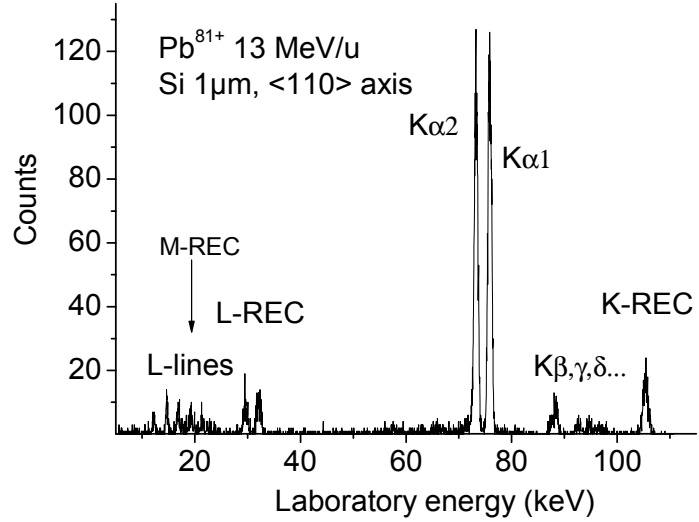

Figure 2: X-ray spectrum detected at 90° for Pb$^{81+}$ incident ions at 13 MeV/u on a 1.1 μm thick silicon crystal, aligned along the <110> direction.

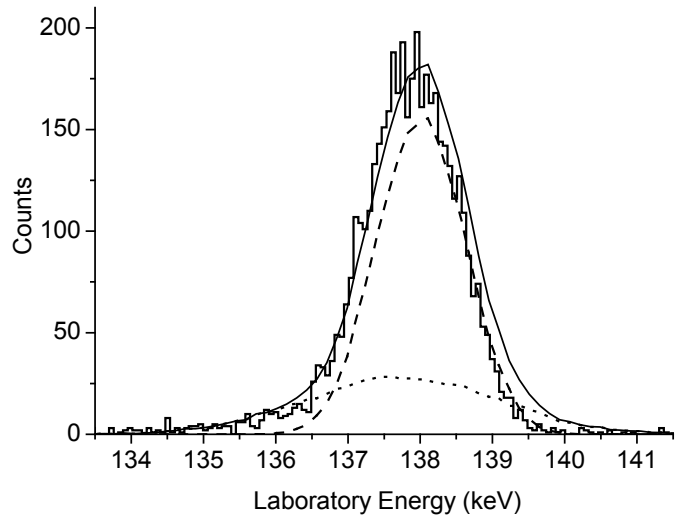

Figure 3: K-REC line shape at 90° in the laboratory frame for 20 MeV/u U$^{91+}$ ions channeled in the <110> axis of a 11 μm silicon crystal. Histogram: experiment. Solid line: simulation. The contributions from the capture of core silicon electrons (dotted line) and valence electrons (dashed line) are shown. The height of the calculated peak has been normalized to the experimental one.